\documentclass[11pt,aps,prb, preprint, showpacs]{revtex4}

\usepackage{amsmath, amsfonts}    
\usepackage{graphicx}   
\usepackage{epstopdf}

\def \be{\begin{equation}}
\def \ee{\end{equation}}

\def \s{{2}^{\frac{1}{12}}}
\def \t{{\mathcal{T }}}
\def \speed{\frac{m}{s}}
\def \min{\emph{minor }}
\def \maj{\emph{major }}
\def \C{$C $}

\newcommand{\singlespace}{
          \renewcommand{\baselinestretch}{1}
           \large\normalsize}
        
\singlespace

\begin{document}

\title{Relativity of musical mood}
\author{Ka\'{c}a Bradonji\'{c}}
 \email{kacha@physics.bu.edu}   
 \affiliation{Boston University, Department of Physics, Boston,
MA 02215}

\date{July 15, 2008}

\begin{abstract}
A combination of three or more tones played together is called a chord. In the chromatic scale, chords which are consonant are of particular interest and can be divided into several groups, two main ones being the major and minor chords. This paper shows that if three sounds are produced by three spatially separated sources, a ``happy" sounding major chord can be observed as its ``sad" sounding counterpart depending on the observer's velocity - a consequence of the well known Doppler effect. The analysis is further extended to show that almost any triad may be observed by choosing an appropriate frame of reference, and several interesting symmetries, asymmetries and features of the system are discussed. Finally, the possibility of applications of this effect in the music performance and creation in the context of ``interactive listener" is discussed and suggestions for overcoming some technical difficulties are proposed.
\end{abstract}


\maketitle

\section{Introduction}
Last 30 years have seen a rich development of the theory of music. Numerous  books and papers have been published aiming to classify musical objects, their relations and transformations, sometimes in very mathematically complex ways. Some of that work focusing specifically on groups of three tones, or triads, deals with what are called \emph{triadic transformation} and is part of a larger \emph{transformational theory}\cite{hook,cohn}. For a simple treatment of some such transformations one can consult Harkleroad\cite{hark}. The aim of this paper is comparatively modest. Although a rich theory of triad transformations has been developed, and the Doppler effect has been taught to high-school students for decades, no one has attempted to put the two together in an applicable way. A lone paper by A. T. Wilson\cite{wilson} relates relativistic rapidity to an analogue of musical pitch in the case of electro-magnetic radiation with a formalism somewhat similar to ours. However, no thorough attempt has been made to investigate the three-dimensional Doppler effect in presence of multiple sources and its application to music. It is known that certain triads sound ``happy", while others sound ``sad". Why this is so has been a question on minds of many musicologists, composers, musicians and music lovers for a very long time and theories have been put forward.  References to some works in this area can be found in the recent book by Loy\cite{loy}. But the question of ``Why?" is beyond the scope of this work. In this paper, we simply show that under specific physical conditions, a chord  sounds happy or sad depending not only on the observer's subjective interpretation, but also on his frame of reference. In other words, the musical ``mood" depends on the observer's state of motion. In the second section, we review sound and the Doppler effect in one and two dimensions. The third section contains an overview of the definitions and properties of tones used in the Western music, (which is the type we are considering), and some general mathematical features of the chromatic scale. Chords and their classifications are also introduced in this section. In section $IV$, we develop a formalism, including a notion of a \emph{triad space} and transformations in this space necessary for our analysis of how the Doppler effect affects the observation of a triad.  Finally, in sections $V$  and $VI$ we show how the ``mood" of the detected chord depends on frame of reference of the observer and how in fact any chord can be detected in an appropriate frame. Some interesting frames of reference as well as their symmetries are also discussed. The paper concludes with a discussion of possibilities this effect opens for musical performance and listening.
\section{Sound and the doppler effect}
Sound waves are essentially pressure waves traveling through an elastic medium. A sound produced by a single source travels in all directions. Point sources produce spherical wavefronts.  A sound wave is characterized by three key quantities: wavelength $\lambda$, frequency $f$, and the speed of propagation $v_{s}$. These three quantities are related by the expression
\be
\label{eq:speed}
v_{s}= \lambda f.
\ee

While the speed and wavelength of a sound wave depend on the medium through which the wave travels, the frequency depends exclusively on the rate of vibration of the source which is producing the sound. 

\begin{figure}[ht]
\begin{center}
\includegraphics[width=1.5in]{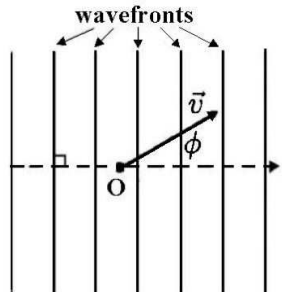}
\caption{\label{fig:kacabradonjicfigure01}Only $v_{o}cos(\phi)$, the velocity component perpendicular to the wavefront, contributes to the Doppler effect in two dimensions.}
\end{center}
\end{figure}

In the late 1842, Christian Doppler\cite{doppler} suggested that the \emph{observed} frequency  of a wave depends on the motion of the source and the observer relative to a stationary medium. This effect, which applies to both sound and light waves, was named the ``Doppler effect" and was soon experimentally confirmed in the case of sound. We imagine that the observer carries  a coordinate system with him wherever he goes, and we call this coordinate system observer's frame of reference. We will consider only inertial frames of reference which are moving at constant velocities. If a sound of frequency $f_{e}$ is emitted by a source moving at speed $v_{e}$ and the observer is moving at speed $v_{o}$, both relative to the stationary medium, the observed frequency $f_{o}$ depends on $v_{e}$ and $v_{o}$ according to\cite{loy}
\be
\label{eq:fulldoppler}
f_{o}=f_{e}\frac{v_{s} + v_{o}}{v_{s} - v_{e}},
\ee 
where $v_{s}$ is speed of sound, and signs of $v_{o}$ and $v_{e}$ both depend on their directions: positive if the observer (source) is approaching the source (observer), and negative otherwise. Note that if $\vec{v}_{o}=\vec{v}_{e}$ and the observer is ``chasing" the source, then the source is  ``running away" from the observer so $v_{o}$ and  $v_{e}$ in Eq.~(\ref{eq:fulldoppler}) have opposite signs. As a result,  the fraction multiplying $f_{e}$ is equal to unity and there is no shift in the observed frequency, just as we would expect. For a source at rest, which is the case we shall consider, $v_{e}=0$ and the Eq.~(\ref{eq:fulldoppler}) becomes
\be
\label{eq:doppler}
f_{o}=f_{e}\frac{v_{s} \pm v_{o}}{v_{s}}.
\ee
The plus (minus) sign is used if the observer is moving towards (away) from the source. This is the one-dimensional form of the Doppler effect - the source is moving along a straight line connecting him to the source. In two dimensions, things are a bit more complicated. If the observer is moving in an arbitrary direction,  only the velocity component that is perpendicular to the wave front of the sound wave will cause a shift in the observed frequency. For a plane wave, such as the one depicted in  Fig.~\ref{fig:kacabradonjicfigure01}, the Doppler effect has form
\be
\label{eq:angle}
f_{o}=f_{e}\frac{v_{s} - v_{o} cos\phi}{v_{s}},
\ee
where $\phi$ is the angle between observer's velocity and the direction of propagation of the sound. If the wave fronts are indeed spherical and the motion is not along a straight line connecting the source and the observer, then the angle will be time dependent. In a case where both the observer and the source are moving, Eq.~(\ref{eq:angle}) has an even more general form. 
\section{Musical math and vocabulary}
Two references that contain thorough reviews of musical objects and their mathematical relations are the famous book by Helmholtz\cite{helm}, as well as the recent book by Loy\cite{loy} and most of what is in this section was found there. Here, we only review what is needed for the purpose of our discussion.
\subsection{Scales and tones}
In the Western common music notation system, a tone is characterized by its pitch, musical loudness, and timbre. While{\bf{frequency}} is a measure of number of vibrations per second, {\bf{pitch}} is the corresponding perceptual experience of frequency\cite{loy}. Since our goal is to show that observed shifts in frequency result in the perception of a modified musical object,  it is crucial for our purpose that frequency and pitch be directly related. So we shall use the two terms interchangeably.  A sequence of pitches and a formula for specifying their frequencies is called a {\bf{scale}}. A scale is usually divided into {\bf{octaves}} which contain specific pitches called {\bf{degrees}}. An {\bf{interval}} is the ratio between frequencies of two tones. We consider the tones of so called {\bf{chromatic}} scale. The chromatic scale consists of tones whose neighboring frequencies differ by a power of $\s$, or a {\bf{semitone}}. The tones of this scale are grouped into octaves of 12 degrees. Frequency $f=440 Hz$ is chosen as a reference, and all the other frequencies are found by multiplying $440 Hz$ by a \emph{power} of $\s$,
\be
f=440\cdot2^{\frac{n}{12}} Hz, \hspace{0.5in} n=0, \pm 1, \pm 2, \pm 3,...
\ee

Tones of the chromatic scale  are labeled by the letters of the alphabet $A$ through $G$ and symbols $\sharp$ and $\flat$.  Symbol ``$\sharp$" denotes a power of $\s$ \emph{increase} in frequency from the letter it follows, while ``$\flat$" indicates a power of $\s$ \emph{decrease} in frequency from the letter it follows. Hence, in this notation $C\sharp$ is the same as $D\flat$.  Octaves of the chromatic scale traditionally begin at $C$, and each tone is labeled with a number that indicates which octave it belongs to. The tones in Table~\ref{tab:kacabradonjictable01} belong to the $4^{th}$ octave and $C$ is more specifically called  $C_{4}$ or {\bf{middle $C$}}. However, as this is the most commonly used octave, its tones are for simplicity denoted without the label ``4". The next octave up begins with  $C_{5}$, and so on. 
\begin{table}[ht]
\begin{center}
\begin{tabular}{ccccccccccccccc}
\hline
\hline
Tone& &\C&  &$C\sharp$&  &$D$& & $D\sharp$& & $E$& &$F$& &$F\sharp$\\
$f(Hz)$& &264.626 & &277.183 & &293.665& &311.127& &329.628& &349.228& &369.994\\
\hline
 Tone& &$G$& &$G\sharp$& &$A$& &$A\sharp$& &$B$& &$C5$&  &$C5\sharp$\\
$f(Hz)$& &391.995& &415.305& &440& &466.164& &493.883& &523.251& &554.365\\
\hline
\hline
\end{tabular}
\caption{\label{tab:kacabradonjictable01}Some tones and their frequencies in the chromatic scale with $f_{A}=440 Hz$ picked as a reference frequency.}
\end{center}
\end{table}

\begin{figure}[ht]
\begin{center}
\includegraphics[width=2in]{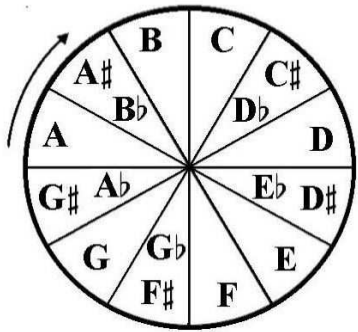}
\caption{\label{fig:kacabradonjicfigure02}The wheel of the chromatic scale is divided into 12 slices, two neighboring slices differing in frequency by a power of $\s$. Chromatic scale progresses in the clockwise direction on the wheel.}
\end{center}
\end{figure}

An alternative representation of the order of tones in the chromatic scale is the commonly used ``wheel" pictured in Fig.~\ref{fig:kacabradonjicfigure02}. One should not be thrown off by the fact that $F$ and $C$ in Fig.~\ref{fig:kacabradonjicfigure02} and Table \ref{tab:kacabradonjictable01} are in the places where one would expect $E\sharp$ and $B\sharp$. This is just a question of notation. The important part is that the neighboring tones of  the chromatic scale are a semitone apart, regardless of their labeling.

Other common kinds of scales are the {\bf{diatonic major}} and {\bf{diatonic minor}}. They both contain eight pitches, and their relation to each other and the chromatic scale within a single octave can be found in Table~\ref{tab:kacabradonjictable02}. 
\begin{table}[ht]
\begin{center}
\begin{tabular}{lllllllllllllll}
 \hline
 \hline
No. of $\frac{1}{2}$-tones above the initial& &0&1&2&3&4&5&6&7&8&9&10&11&12\\
Chromatic& &\C&$C\sharp$& $D$& $D\sharp$& $E$&$F$&$F\sharp$&$G$&$G\sharp$&$A$&$A\sharp$&$B$&$C5$\\
Diatonic $C$ major& &\C&& $D$& & $E$&$F$&&$G$&&$A$&&$B$&$C5$\\
Diatonic $C$  minor& &\C&& $D$& $D\sharp$& &$F$&&$G$&$G\sharp$&&$A\sharp$&&$C5$\\
 \hline
 \hline 
 \end{tabular}
\caption{\label{tab:kacabradonjictable02}The chromatic, diatonic major and diatonic minor scales in the key of $C$.}
\end{center}
\end{table}
A scale is said to be {\bf{in the key}} of the tone it begins with. Table~\ref{tab:kacabradonjictable02} shows the diatonic major and minor scales in the key of $C$, also called the $C$ major and $C$ minor scales. As one can see, the tones of the $C$ major scale are all labeled only by letters, and no $\sharp$'s and $\flat$'s. This is where the somewhat confusing notation in the chromatic scale originates from.
\subsection{Chords}

Three or more tones played simultaneously is called a chord. We will deal with chords consisting of three tones,  or {\bf{triads}}.  The lowest tone in a triad is called a {\bf{bass}} tone. An interval of two pitches is said to be {\bf{consonant}} if the two pitches sound pleasant when played together. Otherwise, it is called{ \bf{dissonant}}. A chord is said to be {\bf{concord}} if the intervals of all the tones in that chord are consonant. There are several types of triads, only two of which we will consider: \maj and $minor$. One should not confuse the major and minor \emph{triads} with the major and minor \emph{scales}, although there is a relation between the two. For that reason, ``major" and ``minor" will be italicized when they refer to a chord, unless it is obvious what is being talked about.

To get a triad within a single octave, one starts from any one tone, called the {\bf{root}} of a triad, and picks two others. To obtain a \maj triad, one starts at the root and picks tones that are 4 and 7 semitones above the root (clockwise in Fig.~\ref{fig:kacabradonjicfigure02}). A \min chord consists of the root,  and tones 3 and 7 semitones above the root. If in the process one reaches the end of the octave and finds the third tone after completing a circle, then the \emph{root} of the triad is not the the same as the \emph{bass} of the triad, and the chord is said to be an {\bf{inversion}}. For instance, $EGC$ and $GCE$ are the inversions of $CEG$ (note the cyclic order.) The chord is named after its root and it is indicated whether it is a \maj  or a \emph{minor}, depending on the separation of its constituent tones (0-4-7 or 0-3-7 semitones apart). Typically, a major chord is labeled simply by the symbol of its root, while a minor chord by the symbol of its root and the letter ``m". Additional notation can be added to indicate whether the chord is an inversion.  

We look at some examples. The chords of the $C$ major scale can be found by consulting Table~\ref{tab:kacabradonjictable02}. Beginning with $C$, we move 4 semitones to the right, where sits  $E$, and then 3 more semitones over to $G$. $CEG$ is a hence a \emph{C major} (simply denoted by $C$). If one starts at $D$, there is nothing 4 semitones to the right, but there is a tone 3 semitones away, namely $F$, followed by $A$ which is 4 semitones further. So we get $DFA$ by picking a tone at 0, 3, and 7 semitones away from D, which makes $DFA$ a \emph{D minor} (or just $D$m). One can continue on like this for all the tones of the $C$ major scale and the same procedure can be done within the $C$ minor scale. Furthermore, one can consider all other major and minor scales, those beginning with $C\sharp$, $D$, $D \sharp$, etc. During this process, one finds that not all the roots will yield a \maj or a \min triad, but that other kinds of triads have to be considered, such as, for example, \emph{diminished} triad whose elements are 0-3-6 semitones apart from the root. A list of triads acquired by this procedure from $C$ major and $C$ minor scales can be found in Table \ref{tab:kacabradonjictable03}.

\begin{table}[ht]
\begin{center}
\begin{tabular}{lcl}
\hline
\hline
$C$ major scale & &$C$ minor scale\\
\hline
$CEG$(\C)& &$CE\flat G$ ($C$m)  \\
$DFA$ ($D$m)& &$DFA\flat$  ($D$ dim.)  \\
$EGB$ ($E$m)& & $D\sharp G A\sharp$ ($D\sharp$)\\
$FAC$ ($F$)& & $F G\sharp C$ ($F$m)\\
$GBD$ ($G$)& & $G A\sharp D$ ($G$m)\\
$ACE$ ($A$m)& & $G\sharp C D\sharp$ ($G\sharp$)\\
$BDF$ ($B$ dim.)& & $A\sharp D F$ ($A\sharp$)\\
\hline
\hline
\end{tabular}
\caption{ \label{tab:kacabradonjictable03}All the triads and inversions found within a single octave of the $C$ major and $C$ minor scales.}
\end{center}
\end{table}
Major and minor chords are all concord. Needless to say, consonance is somewhat of a subjective quality, perhaps rooted in our biology. There are theories on why some combinations of tones sound better, more pleasing, than the others, details of which are beyond the scope of this paper. It is worthy of mention that consonance of two tones seems to be related to the ratio of their frequencies. For a nice physical explanation,  reader can consult Feynman\cite{feynman}. In addition, major chords have been termed happy  and minor chords sad for their perceived emotional character. We will see that in the case where three sound waves are traveling in different directions, the  ``mood" of a consonant chord, and even consonance itself, depends not only on the subjective perception and interpretation of the observer, but also on his motion relative to the source. Next, we develop the formalism necessary for our analysis.
\section{The triad space}
\begin{figure}[ht]
\begin{center}
\includegraphics[width=2.5in]{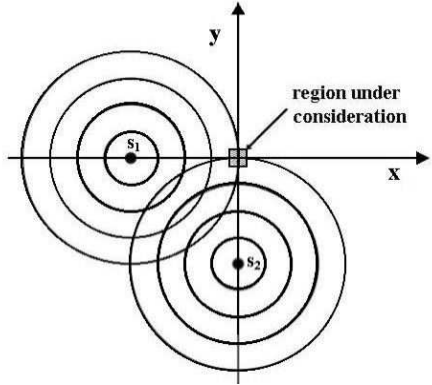}
\caption{\label{fig:kacabradonjicfigure03}In two dimensions, radially propagating waves produced by sources situated on the negative sides of $x$ and $y$ axes far enough from the origin, can be approximated by plane waves in the neighborhood of $(0,0)$. The same approximation can be made in three dimensions, where we add one more source on the negative $z$ axis.}
\end{center}
\end{figure}

Let there be three sources of sound $S_{x}$, $S_{y}$, and $S_{z}$  located on the negative halves of the $x$, $y$, and $z$ axes of the Cartesian coordinate system respectively. We require that the sources are far enough from the origin so that the the three sound waves in the region around $(0,0,0)$ look like plane waves propagating in $\hat{x}$, $\hat{y}$, and $\hat{z}$ directions, in that order. Figure~\ref{fig:kacabradonjicfigure03} shows the region of interest in two dimensions. 

We take the speed of sound to be $v_{s}=343 \frac{m}{s}$, an approximate value of its speed in air under the STP conditions.  An ``observer" $O$ is a person, or a recording instrument, located in the small region near the origin, and capable of detecting sounds coming from \emph{all three} spatial directions. The observer $O$ can be either at rest, or moving relative to the sources with velocity $\vec{v}$. 

Next, we define a three-dimensional {\bf{triad space}} $\t$,  points of which are all possible triads $\bf{f}$ labeled by three frequencies $f_{1}$, $f_{2}$, and  $f_{3}$. Points of $\t$ are denoted by bold letters, while their components with the regular font. Each point $\bf{f} \in \t$ can be represented by a matrix 
\be
\bf{f} =\begin{pmatrix}
f_{1}\\
f_{2}\\
f_{3}\\
\end{pmatrix}.
\ee
Since an observer hears three tones of a triad played simultaneously, he has no way of telling whether he is hearing a particular triad, its inversion, or any other combination of the same three tones. As a result, our formalism allows for multiple matrices representing the same triad and its inversions. The consequences of this multiplicity will become apparent in the upcoming analysis. 

If we pick a single reference triad $\bf{f_{e}}$, then all the other points in $\t$ can be found from the relation 
\be
\label{eq:l}
\bf{f}=\bf{\Lambda f_{e}}=\begin{pmatrix}
2^{\frac{n_{1}}{12}} & 0 & 0 \\
0 & 2^{\frac{n_{2}}{12}} & 0 \\
0 & 0 & 2^{\frac{n_{3}}{12}} \\
\end{pmatrix}\begin{pmatrix}
f_{e1}\\
f_{e2}\\
f_{e3}\\
\end{pmatrix}= \begin{pmatrix}
2^{\frac{n_{1}}{12}} f_{e1}\\
2^{\frac{n_{2}}{12}} f_{e2}\\
2^{\frac{n_{3}}{12}} f_{e3}\\
\end{pmatrix}\hspace{0.5in}n_{i}\in \bf{R},
\ee
where factor of $\s$ was chosen for a later convenience, and $ i=1, 2, 3$ for the rest of the paper. The 3 $\times$ 3 matrix  ${\bf{\Lambda}}$ is a transformation which takes us from the reference triad $\bf{f_{e}}$ to any other point in $\t$. It is completely specified by three values $n_{i}$. Since values of $n_{i}$'s are real numbers, we have a continuous spectrum of triads. 

If we identify the reference triad $(f_{e1},f_{e2}, f_{e3})$ with the triad of tones emitted by the sources $S_{x}$, $S_{y}$, and $S_{z}$, in that order, then taking into account the Doppler effect, the observed triad can presumably be found in $\t$.  Which $\bf{f}$ it will be, depends on the observer's velocity $\vec{v}$. Adapting Eq.~(\ref{eq:doppler}) to three dimensions, we find that the observed triad is related to the emitted triad by
\be
\label{eq:d}
\bf{f}={\bf{D}}\bf{f_{e}}=\begin{pmatrix}1-\frac{\vec{v}\cdot \hat{x}}{v_{s}} & 0 & 0 \\
0 & 1-\frac{\vec{v}\cdot \hat{y}}{v_{s}} & 0  \\
0 & 0 &1-\frac{\vec{v}\cdot \hat{z}}{v_{s}} \\
\end{pmatrix}\begin{pmatrix}
f_{e1}\\
f_{e2}\\
f_{e3}\\
\end{pmatrix}=\begin{pmatrix}
\left(1-\frac{\vec{v}\cdot \hat{x}}{v_{s}} \right)f_{e1}\\
\left(1-\frac{\vec{v}\cdot \hat{y}}{v_{s}} \right)f_{e2}\\
\left(1-\frac{\vec{v}\cdot \hat{z}}{v_{s}} \right)f_{e3}\\
\end{pmatrix}.
\ee
The $3 \times 3$ matrix ${\bf{D}}$ is a transformation specified by three components of the observer $O$'s velocity $\vec{v}$. When acting on the reference triad (which we identified as triad emitted by the sources), ${\bf{D}}$ takes $\bf{f_{e}}$ into a triad whose elements are equal to those observed by $O$. Comparing Eqs.~(\ref{eq:l}) and (\ref{eq:d}), we see that the ${\bf{\Lambda}}$ and ${\bf{D}}$ are equivalent if
\be
\label{eq:velocity}
v_{i}=(1-2^{\frac{n_{i}}{12}})v_{s}.
\ee
The velocity components take on label $i$ whose values $1$, $2$, and $3$ correspond to the $x$, $y$ and $z$ components, in that order.

\begin{figure}[ht]
\begin{center}
\includegraphics[width=4.5in]{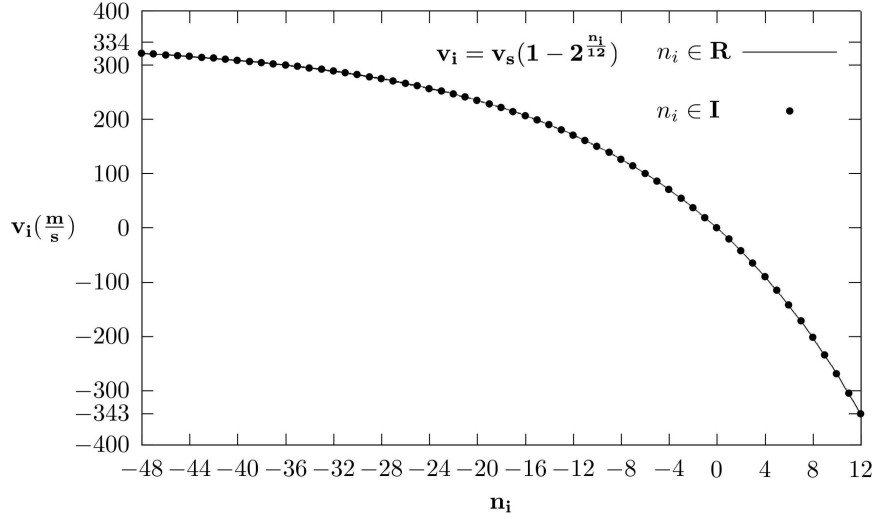}
\caption{\label{fig:kacabradonjicfigure04}Dependence of a velocity component $v_{i}$$\left(\frac{m}{s}\right)$ on parameter $n_{i}$ for both continuous and the discrete spectrum of $n_{i}$. $v_{i}$ asymptotically approaches $v_{s}$ as $n_{i}$ decreases and is equal to $-v_{s}$ at $n_{i}=12$.}
\end{center}
\end{figure}

As it is defined, our space is continuous. But we are only interested in frequencies of the chromatic scale, so we chose $\bf{f_{e}}$ such that its elements are members of this scale. Since the smallest ratio of two frequencies in the chromatic scale is $\s$, only triads we want to consider are those whose elements differ from elements of $\bf{f_{e}}$ by the integer power of $\s$. In our formalism, that is equivalent to restricting $n_{i}$'s  in Eq.~(\ref{eq:l}) to integer values.  This restriction on frequencies also restricts the velocities at which observers can move and still detect frequencies of the chromatic scale. We call these velocities ``allowed" velocities, and they are given by Eq.~(\ref{eq:velocity}), with $n_{i}$'s now restricted to integer values. Equation~(\ref{eq:velocity}) takes care of the direction of the observer's velocity as well. If, for example, $f_{1}>f_{e1}$, then $n_{1}>0$ and we expect $O$ to be moving towards the stationary source (in the negative $x$ direction).  Since $2^{\frac{1}{12}}>1$, we see from Eq.~(\ref{eq:velocity}) that this corresponds to a negative velocity component, just as we expect.  Figure~\ref{fig:kacabradonjicfigure04} shows the plot of $v_i$ vs. ${n_i}$ for both the continuous and the discrete spectrum of $n_{i}$. 
We see that $v_{i}$ asymptotically approaches $v_{s}$ as $n_{i}\rightarrow-\infty$. Curiously enough, going in the other direction $v_{i}$ reaches the speed of sound at exactly $n_{i}=12$, or an octave up. There are physical limitations on the range of the $n_{i}$'s. Only a small range of frequencies is audible to the human ear, and we want to consider only those audible frequencies. Values of $n_{i}$'s depend on these limits, as well as the choice of the reference triad. 
\section{Relativity of the musical consonance and mood}
We begin by showing that a happy triad can be detected as its sad counterpart. Suppose that our three sources are simultaneously emitting $C$, $E$, and $G$, tones of the happy $C$ $major$, so that the reference triad has components with the corresponding frequencies $f_{C}$, $f_{E}$, and $f_{G}$ in that order. Now we ask a question: at what velocity $\vec{v}$ does the observer $O$ have to move so that he detects  $C$, $E\flat$ and $G$, elements of the sad sounding $C$m? 
The simplest answer is found by solving the following equation for $n_{i}$'s and then finding the appropriate velocity components:
\be
\label{majorminor}
\begin{pmatrix}
f_{C}\\
f_{E\flat}\\
f_{G}\\
\end{pmatrix}= \begin{pmatrix}
2^{\frac{n_{1}}{12}} f_{C}\\
 2^{\frac{n_{2}}{12}}f_{E}\\
2^{\frac{n_{3}}{12}} f_{G}\\
\end{pmatrix}.
\ee
We see that $n_{1}$ and $n_{3}$ are obviously zero, as those frequencies remain the same. Since $f_{E\flat}$ is half a tone bellow $f_{E}$, $n_{2}$=-1. Plugging these values into Eq.~(\ref{eq:velocity}), we find that the corresponding velocity is $\vec{v}=19.251 \speed\hat{y}$. So if the sources are emitting $C$, $E$, and $G$, an observer $O_{0}$ situated around the origin and at rest with respect to the origin will detect or hear a \emph{C major}.  An observer $O$ moving with velocity  $\vec{v}=19.251 \speed \hat{y}$ will observe a  \emph{C minor}. This $\vec{v}$, however, is not the only solution. We could also pick an observer $O'$ who is moving with the velocity $\vec{v}'$ leading to the transformation
\be
\label{eq:majorminor2}
\begin{pmatrix}
f_{G}\\
f_{E\flat}\\
f_{C}\\
\end{pmatrix}= \begin{pmatrix}
2^{\frac{n_{1}'}{12}} f_{C}\\
 2^{\frac{n_{2}'}{12}}f_{E}\\
2^{\frac{n_{3}'}{12}} f_{G}\\
\end{pmatrix}.
\ee
Note the swap of $C$ and $G$ on the left hand side. Consulting Table~\ref{tab:kacabradonjictable02}, we see that the solution is $n_{1}'=7$, $n_{2}'=-1$ and $n_{3}'=-7$. Looking at all the combinations which give the elements of $C$m, we see that there are $3!$ in all, equal to the number of permutations of the three values. Table~\ref{tab:kacabradonjictable04} shows all the values  of $n_{i}$'s for which the observer hears the same triad. While the first, the third and the fifth triad in Table~\ref{tab:kacabradonjictable04} are the $C$m and its inversions, the rest are remaining permutations of the tones. If  the observer is capable of distinguishing which direction individual tones are coming from, then he is capable in distinguishing between all six combinations. For simplicity, we assume that is not the case. 

It is interesting to note that the $n_{i}$'s in each row of Table~\ref{tab:kacabradonjictable04} add up to the net number of powers of $\s$ by which the two chords differ. 
\begin{table}[ht]
\begin{center}
\begin{tabular}{ccccccc}
\hline
\hline
Observed triad& &$n_{1}$&$n_{2}$&$n_{3}$& &$n_{1}$+$n_{2}$+$n_{3}$\\ 
\hline
$CE\flat G$& &0&-1&0& &-1\\
$CG E\flat$& &0&3&-4& &-1\\
$E\flat G C $& &3&3&-7& &-1\\
$E\flat C G $& &3&-4&0& &-1\\
$G C E\flat$& &7&-4&-4& &-1\\
$G E\flat C$& &7&-1&-7& &-1\\
\hline
\hline
\end{tabular}
\caption{\label{tab:kacabradonjictable04}Values of $n_{i}$'s for which the same chord is observed if $CEG$ is played, with $f_{e1}=f_{C}$, $f_{e2}=f_{E}$, and $f_{e3}=f_{G}$.}
\end{center}
\end{table}
A similar procedure will show that if \emph{C}m triad is emitted, the observer will detect $C$ \maj if he is in a frame of reference which moves with $\vec{v}=-20.396\speed \hat{y}$ relative to the origin. So it also the case that a sad chord can be detected as its emitted happy counterpart. One can say that there is a relativity of the musical ``mood." 
\section{Observer dependent chords}
\begin{figure}[ht]
\begin{center}
\includegraphics[width=4.5in]{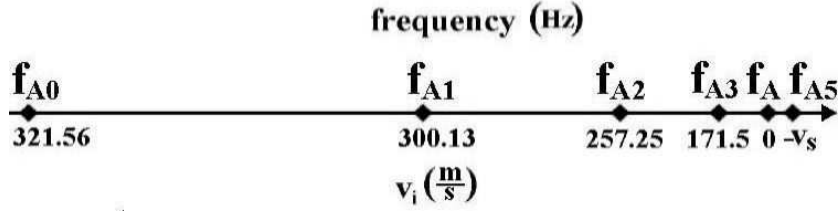}
\caption{\label{fig:kacabradonjicfigure05}Velocity component magnitudes necessary for the observation of a frequency of the chromatic scale for emitted frequency $f_{A}$.}
\end{center}
\end{figure}

It is clear now that, by picking a ``right" frame of reference, any chord can be detected, within the limits imposed on the integers $n_i$. To investigate this possibility further, suppose that the sources are emitting identical tones $A$ of frequency $f_{A}=440 Hz$. Since we are interested in the range of the chromatic scale, and the lower limit of the audible spectrum is around $20 Hz$, we choose the lower limit of the scale at $A0$ of frequency $f_{A0}=27.5 Hz$. This gives us a lower bound on values of $n_{i}$'s of $n_{i(min)}=-48$, corresponding to a shift of four octaves down from $A$. We chose the upper limit to be $C8$ of frequency $f_{C8}=4186.01 Hz$, which is equivalent to the $n_{i(max)}= 87$. However, a problem arises. As it can be seen in Fig.~\ref{fig:kacabradonjicfigure04}, the observer reaches speed of sound at $n_{i}=12$. If we want to avoid sonic booms, we had better limit the positive values of $n_i$'s to 12 or less (even though there may be some issues at $n=12$). This means that possible observable tones will be one ore more octaves bellow and \emph{only} one octave above the emitted frequency, whatever $\bf{f_{e}}$ may be. The asymmetry in the lower and the upper bounds on $n_{i}$'s comes from the power law on which the chromatic scale is built. Figure~\ref{fig:kacabradonjicfigure05} shows how velocities are related to the frequencies of the chromatic scale for the case of $f_{e1}=f_{A}$. 

Values of $n_{i}$'s and the corresponding velocities leading to the observation of the specific tones in this physical situation are listed in Table~\ref{tab:kacabradonjictable05}. 
\begin{table}[ht]
\begin{center}
\begin{tabular}{cccccc|cccccc|cccccc}
 \hline
 \hline
 Tone& &$n_{i}$&   & $v_{i}/v_{s}$&   & Tone& &$n_{i}$&   & $v_{i}/v_{s}$&   & Tone& &$n_{i}$&   & $v_{i}/v_{s}$\\
 \hline
 $A0$& & -48&   &0.938&   &$F$& &-4&   &0.206&   &$C5\sharp$& &5&   &-0.335\\
 $A1$& &-36&   &0.875&   &$F\sharp$& &-3&   &0.159& &$D5\sharp$& &6&   &-0.414\\
 $A2$& &-24&   &0.750&  &$G$& &-2&   &0.109& &$D5$& &7&   &-0.498\\
 $A3$& &-12&   &0.500&   &$G\sharp$& &-1&   &0.056& &$E5$& &8&   &-0.587\\
 \C& &-9&   &0.405&   &$A$& &0&   &0&   &$F5$& &9&   &-0.682\\
 $C\sharp$& &-8&   &0.370&   &$A\sharp$& &1&   &-0.059&   &$F5\sharp$& &10&   &-0.782\\
 $D$& &-7&   &0.333&   &$B$& &2   &&-0.123&  &$G5$& &11&   &-0.888\\
 $D\sharp$& &-6&   &0.293&   &$C5$& &3&   &-0.189&   &$A5$& &12&   &-1\\
 $E$& &-5&   &0.251&   &$C5\sharp$& &4&   &-0.260   & &  &&   && \\
 \hline
 \hline
 \end{tabular}
\caption{\label{tab:kacabradonjictable05}Tones, corresponding $n_{i}$ values, and velocities needed to observe the respective tones when the emitted frequency is  $f_{A}=440 Hz$. The velocities are given in units of speed of sound $v_{s}=343\speed$.}
\end{center}
\end{table}
Since the audibility of a tone depends also on the intensity, and the intensity differs widely for a range of frequencies such as the one we are considering, it is important to assume that the intensity of the emitted sounds is high enough or can be adjusted so that all the tones between $A0$ and $A5$ can, in principle, be either directly heard, or recorded and listened to at some later time.  Also, due to the interference of the three waves, there will be locations where the sounds destructively interfere. We assume that $O$ is located only in those regions where all three tones can be heard.

The list of $n_{i}$'s and $v_{i}$'s leading to observation of various triads is shown in Table~\ref{tab:kacabradonjictable06}. We see that in order for the observer to detect the triad $C$, he has to be moving at the velocity specified by numbers $n_{1}=-9$, $n_{2}=-5$, and $n_{3}=-2$. Similarly, the observer will detect $C$m if in a frame specified by $n_{1}=-9$, $n_{2}=-6$, and $n_{3}=-2$. 
\begin{table}[ht]
\begin{center}
\begin{tabular}{ccccccccc}
\hline
\hline
Triad& &$n_{1}$&$n_{2}$&$n_{3}$& &$v_{1}$ ($\speed$)&$v_{2}$($\speed$)&$v_{3}$($\speed$)\\ 
\hline
$C$& &-9&-5&-2& &139.05&86.04&37.42\\
$C$m& &-9&-6&-2& &139.05&100.46&37.42\\
$F$& &-4&0&-9& &70.761&0&139.05\\
$F$m& &-4&-1&-9& &70.761&19.25&139.05\\ 
$G$& &-2&2&-7& &37.42&-42.00&114.08\\
$G$m& &-2&1&-7& &37.42&-20.40&114.08\\
\hline
\hline
\end{tabular}
\caption{\label{tab:kacabradonjictable06}One of six combinations of $n_{i}$'s and the respective, approximate $v_{i}$'s necessary for the observer to detect the major and minor triads of the $C$ major and $C$ minor scales when all the emitted frequencies are $f_{A}$.}
\end{center}
\end{table}
The multiplicity in possible values of $n_{i}$'s which transform ${\bf{f_{e}}}$ into a same triad is still present, though a bit simpler. Since the three emitted tones have the same frequency, swapping any two values of $n_{i}$, or components of $\vec{v}$, gives the same triad. We still have $3!$ combinations of $n_{i}$'s that lead to the same observation, but now the values of $n_{i}$'s do not change for different combinations - they are only swapped around. In terms of our coordinate system, this means that any transformation of the coordinate axes into each other, while preserving their positive directions, leaves our formalism the same.

An interesting frame of reference is the one whose velocity  $\vec{v}_{T}$ has all components of the same magnitude. In this frame, observer $O_{T}$ detects all the frequencies shifted by the same amount, which is in theory of music known as a $\bf{transposition}$. If numbers $n_{i}$ specifying $\vec{v}_{T}$ are integer multiples of 12,  $n_{i}=12 m$,  $m=0,\pm 1, -2,-3,...$ then the observers in those frames detect triads that are {\bf{octave  equivalent}} of the emitted triad. For example, if $m=-1$, the three emitted $A$'s would be detected as $A3$'s. The trajectory of such an observer is any path parallel to a line described by 
\be
\vec{r}=t(\hat{x}+\hat{y}+\hat{z}), \hspace{0.5in}\mbox{for } -\infty<t<\infty,
\ee
so as long as it remains in the region in which the plane wave approximation holds. (Note that due to our limitation on $n_{i}$'s, one would need emitted frequencies to be higher than $A$ in order to be able to detect anything higher than $A5$.)

At first sight it may seem that \emph{adding} $\vec{v}_{T}$ to the velocity $\vec{v}$ of an observer would result in $O$ detecting the allowed transposition of the triad detected at $\vec{v}$. That, however is not the case. Since $v_{i}$'s depend non-linearly on $n_{i}$'s, adding two allowed velocities does not in general result in an allowed velocity because
\be
\label{eq:sum}
v_{i}+v_{i}'=v_{s}(1-2^{\frac{n_{i}}{12}}+1-2^{\frac{n_{i}'}{12}})\neq v_{s}(1-2^{\frac{m_{i}}{12}}) \mbox{, for }m_{i}=0,\pm1, \pm2,...
\ee
Components of a velocity that  \emph{can}  be {\emph{\bf{added}}} to $\vec{v}$ in any one, or all directions, are
\be
u_{i}=(1-2^{\frac{p_{i}}{12}})2^{\frac{n_{i}}{12}}, \hspace{0.5in} \mbox{for } -(48+n_{i})\leq p_{i} \leq (12-n_{i}),
\ee
where given lower limits on $p_{i}$'s apply to the case when the elements of ${\bf{f_{e}}}$ are all $f_{A}$. In general, they depend on the emitted frequencies and the audibility threshold. In a vector notation, we have
\be
\label{eq:u}
\vec{u}=(1-2^{\frac{p_{1}}{12}})(2^{\frac{n_{1}}{12}})\hat{x}+(1-2^{\frac{p_{2}}{12}})(2^{\frac{n_{2}}{12}})\hat{y}+(1-2^{\frac{p_{3}}{12}})(2^{\frac{n_{3}}{12}})\hat{z}.
\ee
One can check that a sum of any velocity of form specified in Eq.~(\ref{eq:velocity}) (with $n_{i}$'s restricted to integers) and $\vec{u}$ is itself an allowed velocity. Note that what this operation essentially does is ``cancel" the effects specified by $n_{i}$ and replaces them with effects specified by $(p_{i}+n_{i})$. Depending on $n_{i}$, given value of $p_{i}$ will either increase or decrease the observed $f_{i}$, so there is \emph{no need to subtract} $u_{i}$ from $v_{i}$. In fact, if we did that, we would find that it can be meaningfully done only in the case of $p_{i}=12$ in which case the frequency is shifted up by an octave and $\vec{v}'=-v_{s}$. The value $p_{i}=0$ corresponds to adding a zero velocity. 

If an observer $O$ moving with  velocity $\vec{v}$ observes a triad $\bf{f}$, then the observer $O'$ moving at $\vec{v}'=\vec{v}+\vec{u}$ will observe $\bf{f'}$ that is a  \emph{transposition} of $\bf{f}$ by $p$ semitones  \emph{if} $p_{1}=p_{2}=p_{3}=p$. In the case of $p_{1}+n_{1}=p_{2}+n_{2}=p_{3}+n_{3}=l$, the observed triad will be a transposition of the \emph{emitted} triad $\bf{f_{e}}$ by $l$ semitones. Otherwise, the addition will result in the observation of a non-transposition of either $\bf{f}$ or  $\bf{f_{e}}$, and its elements will depend on $p_{i}$'s.
\section{Conclusions and some thoughts on a creative observer}
We saw that if three tones of a major chord are played from three different locations, the perception or observation of the chord depends on the observer's frame of reference. In order that frequencies of the chromatic scale be observed, certain restrictions were placed on observer's velocity and a framework within which one can determine what velocities are allowed was developed. We have shown that there is sort of a relativity of the musical mood and how to go from a happy reference frame to a sad one by choosing appropriate velocity for the observer. The physical situation described in section $IV$ and the subsequent analysis is a rough approximation and it holds only in the region where the three sound waves can be approximated by plane waves traveling in $\hat{x}$, $\hat{y}$, and $\hat{z}$ directions. A more general analysis could be done where one takes into account the spherical nature of the waves. In this case, the observer is free to move anywhere in space, and allowed velocities are position dependent and determined by the location of the sources.  

Besides being a neat application of the Doppler effect, the triad dependence on the observer's frame of reference yields some interesting possibilities in the realm of musical performance and listening. Doppler shifts have been used in practice by creating musical pieces with sound-emitting mobile phones  swung on strings\cite{phone}. The Doppler effect was also noted in a massive ``concert on bicycles" where an audience of riding cyclists was listening to a radio broadcast from units positioned on  moving bikes\cite{schiemer2}. Although these cases involved a moving source rather than a moving observer, the principle is the same.

Idea of interactive music in video games, where the audio content of a game depends on the actions of the player has drawn a lot of attention in recent years and some work has been done on the subject\cite{interaction1}. Reference frame dependence of the triad can take the possibility of the interactive listener out of the virtual world and into the physical reality. Multiple projects in the realm of interactive observer/listener/composer are being conducted at the MIT Media Lab by Machover\cite{Machover}, Paradiso\cite{paradiso1,paradiso2}, as well as others\cite{blockjam}. Our analysis deals with observers in different reference frames. However, if the observer were allowed to change his velocity and was able to do that in a short enough time, he would observe a sequence of shifts of the emitted triads, hence actively participating in his listening experience of a simple chord or a more complex musical piece.  A technical problem with the applicability of the triad frame dependence arises due to the high velocities necessary to achieve a wide enough spectrum of observed sounds. One alternative is to leave the listener at rest, but give him a control over the motion of the sources along the axes on which they are situated. However, as it can be seen from Eq.~(\ref{eq:fulldoppler}), this would require a different formalism than the one presented because the Doppler effect for sound is different depending on whether the observer is moving while the source is at rest, or it is the other way around\cite{mathpages}.

Also, an intermediary recording instrument capable of moving at high speeds could be constructed so that it can be controlled by the listener. The listener could then change the motion of the intermediary device in real time, hence actively participating in his listening experience. Another interesting option is to place the intermediary device in an isolated environment in which density, temperature and pressure of air could be adjusted during the listening so that the speed of sound, and hence all the allowed speeds, would be lower.

Since the Doppler effects have been used in the computer generated music, the first step in development of something of this kind would be a creation of a computer simulation which models the situation we analyzed using the presented parameters, and enables the listener to control the motion of an intermediary virtual device relative to the virtual sources.

As for the application to the musical composition, one could write ``path dependent" pieces, experience of which varies with the composer-prescribed paths that are to be traveled by the observer who is, over the course of the performance, moving at non-zero velocity and occasionally briefly accelerates in order to switch to different reference frames. 

\begin{acknowledgments}
I'd like to thank Kipton Barros and John Swain for reading the earlier drafts of this paper, as well as their insightful comments.
\end{acknowledgments}


\begin{thebibliography}{5}

\bibitem{hook}Julian Hook, ``Uniform Triadic Transformations," J. of Music Theory,  {\bf{46}}(1), 57-126 (2002).
\bibitem{cohn}Richard Cohn, ``Neo-Riemannian Operations, Parsimonious Trichords, and Their `Tonnetz' Representations," J. of Music Theory, {\bf{41}}(1), 1-66 (1997).
\bibitem{hark} Leon Harkleroad, \textsl{The Math Behind the Music} (Cambridge University Press, New York, NY, 2007), p.33.
\bibitem{wilson}Alma T. Wilson, ``Relativistic rapidity as change in musical pitch," \\
\url{<http://lanl.arxiv.org/pdf/0706.3247>}.
\bibitem{loy} Gareth Loy,  \textsl{Musimathics: The Mathematical Foundations of Music, Volume 1}, (MIT Press: Cambridge, MA, 2006).
\bibitem{doppler} Peter M. Schuster, \textsl{Moving the Stars: His Life, His Works and Principle; Christian Doppler and the World After}, Tran. by Wilmes, L. (P\"{o}llauberg, Austria: Living Edition, 2005).
\bibitem{helm} Hermann Helmholtz,  \textsl{On the Sensations of Tone as a Physiological Basis for the Theory of Music} (Dover Publications, New York, NY 1954).
\bibitem{feynman} Richard P. Feynman,  Robert B. Leighton  and Matthew L. Sands, \textsl{The Feynman Lectures of Physics, Volume 1}, (Addison-Wesley Publishing Company: Reading, MA, 1989).
\bibitem{phone}Greg Schiemer and Mark Havryliv, ``Pocket  Gamelan: interactive mobile music performance," in \textsl{Proceedings of Mobility Conference 2007: The 4th International Conference on Mobile Technology, Applications and Systems: IS-CHI 2007}, edited by A.D. Cheok,  P. H. Chong,  W. Sheah, and S. Ping, S. (Research Publishing, Singapore, 2007), p. 716-719. 
\bibitem{schiemer2}Greg Schiemer,  ``Interactive Radio," Leonardo Music Journal, {\bf{ 4}}, 17-22 (1994).
\bibitem{interaction1} David Lieberman, ``Game enhanced music manuscript," in \textsl{Proceedings of the 4th international Conference on Computer Graphics and interactive Techniques in Australasia and Southeast Asia GRAPHITE `06}(ACM, New York, NY, 2006), p. 245-498.
\bibitem{Machover} Research web page of Tod Machover, \url{<http://www.media.mit.edu/research/35>}.
\bibitem{paradiso1}Mark Feldmeier and Joseph A. Paradiso, ``An Interactive Environment For Large Groups With Giveaway Wireless Motion Sensors," Computer Music Journal, {\bf{31}}(1),  68-93 (2007).
\bibitem{paradiso2} Research web page of Joseph A. Paradiso, \url{<http://www.media.mit.edu/resenv/>}.
\bibitem{blockjam} Henry Newton-Dunn, Hiroaki Nakano, and James Gibson, ``Block Jam: A Tangible Interface for Interactive Music"  \textsl{Proceedings of the 2003 Conference on New Interfaces for Musical Expression (NIME-03), Montreal, Canada}, 170-177.
\bibitem{mathpages} Kevin Brown, \textsl{Reflections on Relativity (Mathpages.com, 1999)},  WWW Document, \\ \url{<http://www.mathpages.com/rr/rrtoc.htm>}.
\end{thebibliography}
\end{document}